# Immunohistochemistry guided segmentation of benign epithelial cells, *in situ* lesions, and invasive epithelial cells in breast cancer slides

Authors: Maren Høibø[1,2*], André Pedersen[1,3,4,5], Vibeke Grotnes Dale[1,6], Sissel Marie Berget[6], Borgny Ytterhus[1], Cecilia Lindskog[7], Elisabeth Wik[8,9], Lars A. Akslen[8,9], Ingerid Reinertsen[4,10], Erik Smistad[4,10], Marit Valla[1,2,3,6]

1. Department of Clinical and Molecular Medicine, Norwegian University of Science and Technology (NTNU), Trondheim, Norway
2. Clinic of Laboratory Medicine, St. Olavs hospital, Trondheim University Hospital, Trondheim, Norway
3. Clinic of Surgery, St. Olavs hospital, Trondheim University Hospital, Trondheim, Norway
4. Department of Health Research, SINTEF Digital, Trondheim, Norway
5. Application Solutions, Sopra Steria, Trondheim, Norway
6. Department of Pathology, St. Olavs hospital, Trondheim University Hospital, Trondheim, Norway
7. Department of Immunology, Genetics and Pathology, Cancer Precision Medicine Research unit, Uppsala University SE-751 85 Uppsala, Sweden
8. Centre for Cancer Biomarkers (CCBIO), Department of Clinical Medicine, University of Bergen, Bergen, Norway
9. Department of Pathology, Haukeland University Hospital, Bergen, Norway
10. Department of Circulation and Medical Imaging, Norwegian University of Science and Technology (NTNU), Trondheim, Norway

*maren.hoibo@ntnu.no

## Abstract

**Background:** Digital pathology enables automatic analysis of histopathological sections using artificial intelligence (AI). Automatic evaluation could improve diagnostic efficiency and help find associations between morphological features and clinical outcome. For development of such prediction models, identifying invasive epithelial cells, and separating these from benign epithelial cells and *in situ* lesions would be the first step. In this study, we aimed to develop an AI model for segmentation of epithelial cells (benign, *in situ*, and invasive) in sections from breast cancer.

**Methods:** We generated epithelial ground truth masks by restaining hematoxylin and eosin (HE) sections with cytokeratin (CK) AE1/AE3, and by pathologists' annotations. HE/CK image pairs were used to train a convolutional neural network, and data augmentation was used to make the model more robust. Tissue microarrays (TMAs) from 839 patients, and whole slide images from two patients were used for training and evaluation of the models. The sections were derived from four cohorts of breast cancer patients. TMAs from 21 patients from a fifth cohort was used as a second test set.

**Results:** In quantitative evaluation, a mean Dice score of 0.70, 0.79, and 0.75 for invasive epithelial cells, benign epithelial cells, and *in situ* lesions, respectively, were achieved. In qualitative scoring (0-5) by pathologists, results were best for all epithelium and invasive epithelium, with scores of 4.7 and 4.4, respectively. Scores for benign epithelium and *in situ* lesions were 3.7 and 2.0, respectively. The model is made freely available in the software FastPathology and the project code is available at https://github.com/AICAN-Research/breast-epithelium-segmentation

**Conclusion:** The proposed model segmented epithelial cells in HE stained breast cancer slides well, but further work is needed for accurate division between benign, *in situ,* and invasive cells. Immunohistochemistry, together with pathologists' annotations, enabled the creation of accurate ground truths.

# Background

Most pathology laboratories are burdened by an increased workload. With an aging population and rising cancer incidence (1), the number of biopsies is increasing. Advances within molecular pathology have resulted in more complex diagnostics, expanding the workload for each biopsy. The recent and ongoing implementation of digital pathology makes it possible to analyze tissue sections on computer screens, and it facilitates distant collaboration and the use of digital microscopy in teaching (2). Artificial intelligence (AI) opens for automatic interpretation of digital tissue slides, with the potential to improve diagnostic efficiency (3) and discover novel features of clinical importance. In pathology, AI has shown promising results in tasks like tissue segmentation, mitosis detection, and prediction of prognosis (4-6). Within AI, deep learning has become the preferred method for image analysis. In segmentation of medical images, U-Nets (7) are widely used (8), usually with supervised learning through annotated data. Attention-gated U-Nets with multiscale input and deep supervision are shown to outperform regular U-Nets (9). In semantic segmentation tasks, each pixel in the training images has a label assigned to it, and the neural network will try to learn to label each pixel correctly. The large size of whole slide images (WSIs) represents a significant challenge in image analysis. They can be as large as 200 000 x 100 000 pixels with multiple image planes. This results in a need for large storage capacity, long processing time and complicated data handling.

Methods that enable automatic detection of invasive epithelial cells could be useful in numerous tasks, such as automatic identification of lymph node metastases, automatic biomarker assessment, and prediction of prognosis. Epithelial cells can be detected through immunohistochemical staining with cytokeratins (CK), such as CK AE1/AE3 (10, 11). However, the marker does not differentiate between benign epithelial cells, non-invasive *in situ* lesions, and invasive epithelial cells. In tasks such as automatic biomarker assessment, separating these cells is important, since only invasive cells are included in the analysis.

Segmentation of the tumor border using manual annotations as ground truth has been done previously (12). Manual annotations have also been used to segment invasive, benign, and *in situ* lesions in multiclass models (13-15). However, accurate manual annotations for separating invasive epithelial cells from other cell types within the tumor region would be too time-consuming on large datasets and therefore not feasible. IHC generated ground truths could therefore be an alternative. Bulten *et al*. (16) segmented epithelial cells in HE stained prostate cancer slides, using CK to detect epithelial cells, and a myoepithelial cell marker to separate benign and invasive cells. Brázdil *et al.* (17) used CK to distinguish epithelial cells from surrounding stromal tissue in sections from breast and colon cancer. They did not differentiate between neoplastic and non-neoplastic epithelium. In automatic Ki-67, estrogen receptor (ER), and progesterone receptor (PR) analysis, Valkonen *et al.* (18) segmented breast cancer epithelial cells using a pan-cytokeratin antibody. Their model did not differentiate between invasive and non-invasive epithelium. Although the use of IHC for segmentation tasks shows promise, separating invasive epithelial cells from benign epithelium and *in situ* lesions still remains a challenge that needs to be solved.

The aim of this study was to construct an AI model for segmentation of benign, *in situ*, and invasive epithelial cells in HE stained breast cancer sections, using HE and IHC image pairs and pathologists' annotations to create ground truth.

The main contributions of this paper are: 1) A novel breast cancer dataset comprising HE and CK TMA image pairs from 860 patients, as well as whole slide images from two breast cancer patients. All sections include pathologists' annotations of benign epithelium and *in situ* lesions; 2) An algorithm for extracting TMA cores from histopathological images (TissueMicroArrayExtractor in pyFAST (19, 20)); 3) An algorithm for creating ground truths for HE images based on CK stained images; 4) A trained multiclass segmentation attention-gated U-Net model separating epithelium into benign, *in situ*, and invasive; 5) Comprehensive quantitative and qualitative validation studies;

and 6) The model is made available in the open software FastPathology (21), and the project code is available at https://github.com/AICAN-Research/breast-epithelium-segmentation

## Materials and methods:

### Cohorts and tissue specimens

In this study, TMAs from five cohorts (22-26) of breast cancer patients were used.

- BCS-1 comprises 909 women diagnosed with breast cancer from a background population of 25 727 women, born between 1886-1928 in Trøndelag County, Norway, who were invited to participate in a population-based survey (22) for the early detection of breast cancer. They were followed for breast cancer occurrence from January 1$^{st}$, 1961, to December 31$^{st}$, 2008.
- BCS-2 comprises 514 women diagnosed with breast cancer from a background population of 34 221 women, born between 1897-1977 in Trøndelag County, Norway, who participated in a population-based survey (23). They were followed for breast cancer occurrence from attendance (1995-1997) to December 31$^{st}$, 2009.
- BCS-3 comprises 533 women diagnosed with breast cancer from a background population of 22 941 women born between 1920 and 1966 at EC Dahl's Foundation in Trondheim, Norway. They were followed for breast cancer occurrence from 1961 to 2012 (24).
- HUS-BC-1 comprises 534 women from Hordaland County, Norway, diagnosed with breast cancer from 1996-2003. The patients in this cohort were diagnosed through the national breast cancer screening program and were in the age range 50-69 years (25).
- HPA-BC comprises TMA sections from 25 patients from the Uppsala Biobank/Human Protein Atlas (26).

In this study, four TMA slides from BCS-1, BCS-2, and BCS-3, twelve TMA slides from HUS-BC-1 and four TMA slides from the HPA-BC cohorts were used. In addition, two WSIs from BCS-2 were included. The TMA slides from BCS-1, BSC-2, BCS-3, and HUS-BC-1 comprised 2925 TMA cores from 967 patients. 709 TMA cores were excluded, either because they were missing, displaced, improperly CK stained or due to broken tissue between stains. After these exclusions 2 216 cores from 839 patients were included in the study. Test set 2 comprised 96 TMA cores from 25 patients. After exclusion 56 cores from 21 patients were included in the study.

The TMA slides from BCS-1, BCS2, BCS-3, and HPA-BC were 4 µm thick with a core diameter of 1 mm. The slides from HUS-BC-1 were 5 µm thick with core diameters of 0.6 and 1 mm. All TMA slides were stained as follows: The slides were placed in TissueClear (Sakura Finetek Norway) for 2x3 minutes and rehydrated in four decreasing ethanol dilutes {100, 100, 96, 80} % for 35 seconds each, before being placed in water for 1 minute. The slides were then stained with hematoxylin for 5 minutes, followed by 5 minutes in water, and 1 minute in 80 % ethanol. The slides were stained with eosin (alcoholic) for 1 minute, followed by placement in {96, 96, 100, 100, 100} % ethanol for {30, 30, 60, 60, 60} seconds, respectively. Finally, the slides were placed in TissueClear, first for 30, then for 40 seconds, before being air-dried for 2 minutes. The HE stained slides were scanned using Olympus BX61VS with VS120S5 at x40 magnification and extended focal imaging (EFI) with seven planes. The scanned HE slides were inspected by a pathologist. The coverslips were then removed from the HE stained slides by placement in Xylene for 2-3 days, then placed in TissueClear for 3x2 minutes, and rehydrated in decreasing ethanol dilutes: 3x2 minutes in 100 %, 1x2 minutes in 96 %, 1x2 minutes in 80 %, and 2x5 minutes in water. Antigen retrieval was performed through heat induced epitope retrieval (HIER) by placing the slides in DAKO/Agilent TRS-buffer pH 9 (Agilent DAKO, Agilent Technologies Denmark ApS) for 20 minutes at 98 °C and 2x5 minutes in DAKO wash buffer (Agilent DAKO, Agilent Technologies Denmark ApS). The HE stain was removed during the HIER procedure. The slides were restained with the CK AE1/AE3 antibody (Agilent DAKO, Agilent Technologies Denmark ApS) through four incubation steps: five minutes incubation in hydrogen peroxide, 90 minutes incubation with primary antibody for pan CK AE1/AE3 (concentration 176.7 mg/L, dilution:

1:80), 30 minutes incubation with EnVision K5007 rabbit/mouse Horseradish Peroxidase (HRP)/Diaminobenzidine (DAB)+ detection Kit (Agilent DAKO, Agilent Technologies Denmark ApS), and 2x5 minutes incubation with DAB+ chromogen. DAKO wash buffer was used between each incubation step. The slides were then stained with contrasting hematoxylin for 15 seconds, dehydrated, and coverslips were applied. Dehydration was performed through placement in ethanol (1 minute in 80 %, 1 minute in 96 %, and 3x1 minute in 100 %). The CK AE1/AE3 stained slides were scanned with Olympus BX61VS with VS120S5 at x40 magnification using EFI with seven image planes.

The two WSIs were stained with HE and scanned using Olympus BX61VS with VS120S5 at x40 magnification without EFI, then CK stained and rescanned. The staining procedures were identical to those for the TMAs.

Annotations

The stains in the CK images were separated automatically in QuPath (27) using color deconvolution (28) by setting the image type to brightfield Hematoxylin-3,3'-Diaminobenzidine (H-DAB). This produced artificial deconstructed Hematoxylin, DAB, and residual stains. The default stain vector in QuPath for one of the slides was used as reference for all slides. Preliminary ground truth masks were created by thresholding the DAB stain channel using a Gaussian prefilter, smoothing sigma 3.0, and a threshold of 0.25 at 0.3448 µm/pixel resolution (x20 magnification). Since CK is a cytoplasmic marker, nuclear holes were present in the mask. Holes with an area below 150 µm$^2$ were therefore filled, and small fragments were discarded by removing objects with an area smaller than 25 µm$^2$ (see Figure 1c). The masks were exported from QuPath to GeoJSON then converted to tiled, non-pyramidal TIFF followed by a conversion to pyramidal BigTIFF. Since CK, which in the image corresponds to the DAB stain, stains all epithelium, the initial mask consisted of only two classes: epithelium and non-epithelium (see Figure 1d).

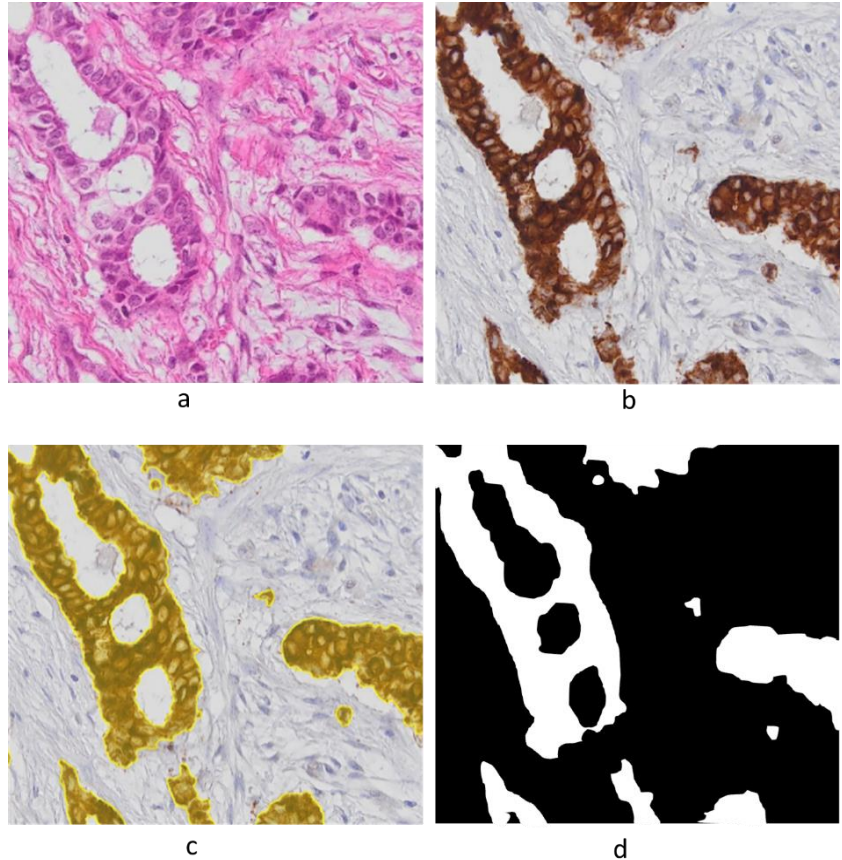

a  b

c  d

Figure 1: a) HE stained tissue. b) CK stained tissue. c) Thresholded DAB channel in CK stained tissue where small holes are filled, and fragments removed. d) Illustration of binarized thresholded DAB channel. Abbreviations: HE = hematoxylin and eosin, CK = cytokeratin, DAB = 3,3'-diaminobenzidin.

A pathologist reviewed the CK images and marked sections with strong background staining to avoid false positives and sections with false negative CK staining. TMA cores with more than 10 % false negative epithelium staining or strong background staining were tagged in QuPath for exclusion. The HE images were annotated by two pathologists, who digitally marked benign epithelial structures and all *in situ* lesions (including all non-invasive atypical epithelial cell proliferations) using QuPath to separate these from invasive epithelial cells (see Figure 2d). The pathologists reviewed each other's annotations in the HE images. Consensus was reached through discussion in case of discrepancies.

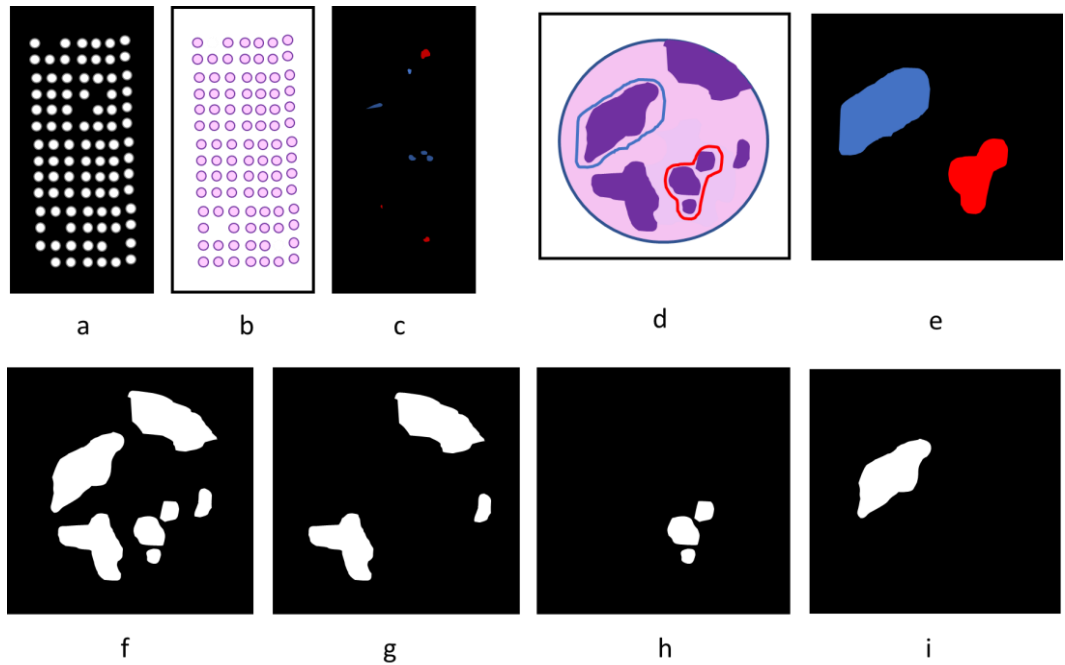

Figure 2: Illustrations of a) TMA slide of binarized DAB channel. b) HE stained TMA slide. c) TMA slide of manual annotations of benign epithelium and *in situ* lesions. d) HE stained TMA core with manual annotations of benign epithelium and *in situ* lesions. e) Manual annotations of benign epithelium and *in situ* lesions in TMA core. f) Binarized DAB channel, positive for all epithelium. g) Invasive epithelium. h) Benign epithelium. i) *in situ* lesion. Abbreviations: TMA = Tissue microarray, DAB = 3,3'-Diaminobenzidin, HE = Hematoxylin and eosin.

To enable evaluation of TMA cores according to histological subtype and grade, each case (TMA triplet or duplet) was identified and marked in QuPath. The annotations of benign epithelial structures and *in situ* lesions, annotations of cores with insufficient staining, and case annotations were exported from QuPath to OME-TIFF format as three separate images.

### Dataset creation
#### Tissue Microarrays
The HE and CK images, the preliminary DAB ground truth mask, annotations of benign epithelium and *in situ* lesions, annotations of cores to be removed, and case annotations were imported into Python as six separate pyramidal images.

Since multiple TMA cores existed for each patient, the data was divided into training, validation, and test sets on slide level. This ensured that the same patient was not present in multiple data sets.

The slides from the BCS-1, BCS-2, BCS-3, and HUS-BC-1 cohorts were randomly divided into training, validation, and test sets, where 16, 4, and 4 slides were used in training, validation, and test sets, respectively. The HPA-BC cohort was used exclusively as a second test set.

An image processing algorithm (called TissueMicroArrayExtractor) for extracting TMA cores automatically from TMA whole slide images was developed and added to the open-source library FAST (19, 20) and the corresponding Python interface pyFAST. This algorithm first performs tissue segmentation by color thresholding on a low-resolution version of the image. The segmentation regions are then extracted using flood fill. Small regions of less than 100 pixels are removed, and for the rest, the median area and diameter is calculated. Any region that differs more than 50% from the median area or diameter are excluded. The final regions are then extracted from the desired magnification level. The TMA cores were extracted automatically from the HE and CK slides at x10 magnification from the pyramidal images using the TissueMicroArrayExtractor, to get sufficient details while still including larger tissue structures within a patch. HE/CK core pairs were identified by comparing coordinates of the extracted cores. Corresponding areas were extracted from the annotated and thresholded images using the coordinates, and width/height of the HE/CK cores. Cores marked as insufficiently CK stained were not included in the analysis, as well as cores that were severely displaced or destroyed. Smaller displacements were identified and adjusted for with registration of the CK and HE cores using phase cross-correlation. Histogram equalization was used to increase the contrast in the CK cores to improve registration of tissue with few details and poor contrast. The TMA core images were then downsampled with a factor of four to reduce memory use during registration. The shifts in x and y-directions between the HE and CK cores were calculated, before being upscaled and used to register the images at full resolution.

Invasive epithelium, benign epithelium, and *in situ* lesions were separated into three separate classes for each TMA core (see Figure 2f-i). Ground truth for invasive epithelium was created by subtracting CK stained epithelium within areas manually annotated as benign or *in situ* from the mask made by thresholding the DAB channel. Benign epithelium and *in situ* lesion ground truths were created by identifying the positive cells from the mask (see Figure 2a and f) within areas annotated as benign or *in situ* by the pathologists (see Figure 2c, e, h, i).

Due to the large size of histopathological images, patches of size 1024 x 1024 pixels were created from the HE, CK, benign, *in situ*, and invasive TMA core images with 25 % overlap on all sides using pyFAST. Patches with less than 25 % tissue were excluded from training. To improve registration, the HE and CK patches were registered again, as described for the whole TMA cores earlier. A ground truth patch was created by one-hot encoding the non-epithelial tissue, invasive epithelium, benign epithelium, and *in situ* lesions as four separate classes. Due to the large imbalance between the number of patches including invasive epithelium versus benign and *in situ*, the patches were divided into benign, *in situ*, and invasive sets, allowing for a balanced sampling scheme during training. For each patch, the patch was assigned to the *in situ* set if the patch included *in situ* lesions. If the patch included benign epithelium but not *in situ* lesions, it was assigned to the benign set. If the patch did not include *in situ* or benign epithelium, the patch was assigned to the invasive set (see Figure 3).

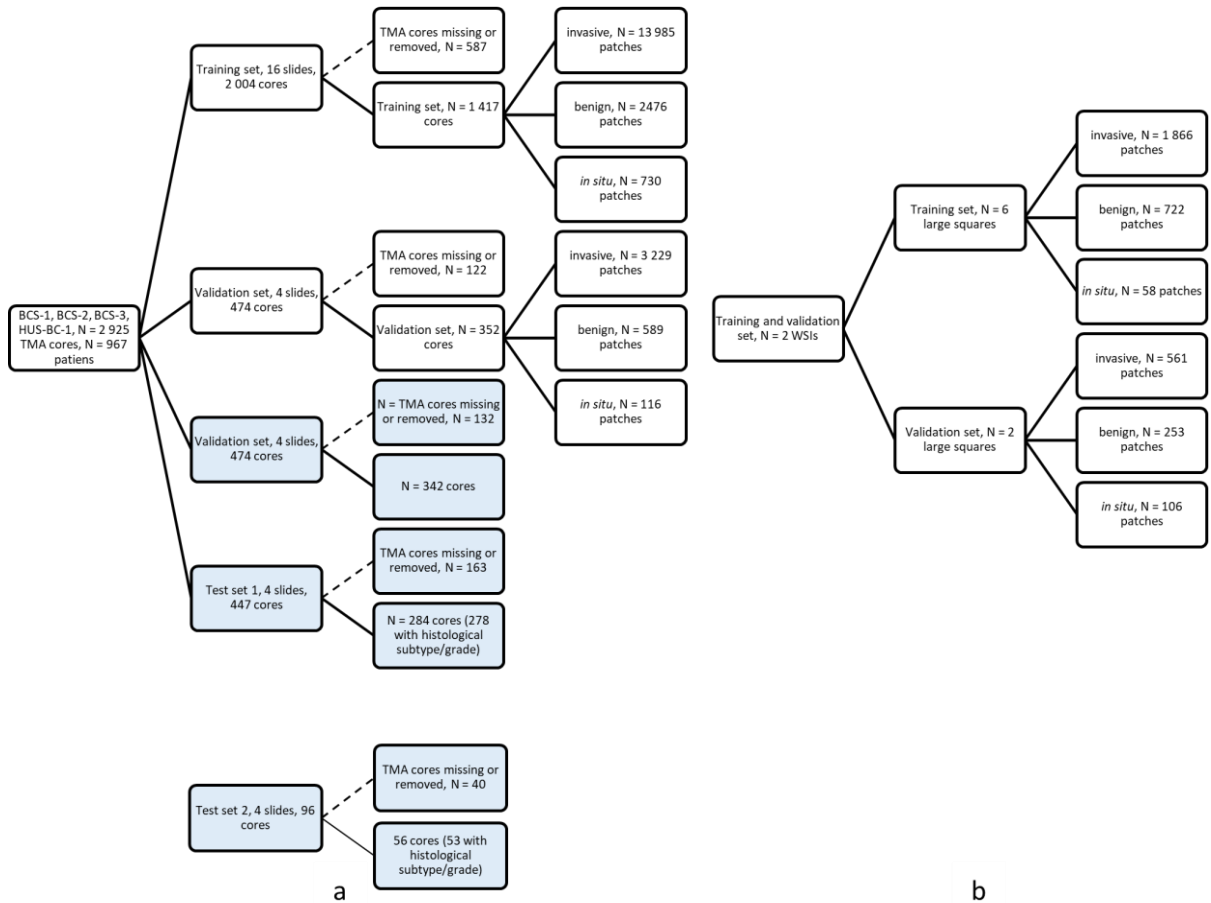

Figure 3: Dataset creation. a) Data stratification from TMAs. Blue boxes represent whole TMA cores (image level 1). White boxes represent training and evaluation data (patches extracted at image level 2). b) Data stratification from WSIs. Abbreviations: TMA = Tissue microarray, WSI = Whole slide image.

Whole slide images
Selected areas from two whole sections of breast cancer were added to the training data to include more tissue from areas poorly represented in TMAs, such as benign epithelium, *in situ* lesions, stromal tissue, and adipose tissue. Areas with only invasive epithelial cells or glass were annotated for removal, as well as some regions including large amounts of adipose tissue. The two slides were used for training and validation. The WSIs were annotated for benign epithelium and *in situ* lesions and reviewed for false positive and false negative CK staining, as previously described. The WSIs were first divided into eight squares, which were registered using phase cross-correlation (downsampled, registered, and shifted). Six of the large squares were included in the training set and two in the validation set. Patches from the same slide could thus be found in both the training and validation set. Creation of HE patches and corresponding ground truths were performed as described for the TMAs.

Training and evaluation
To counter class imbalance, patches were randomly sampled during training from one of the three sets (benign, *in situ*, or invasive) from the TMAs and WSIs. During patch generation, data augmentation was performed concurrently to the training data to improve model robustness. The following data augmentation techniques were used: random flip, 90° rotations, brightness, hue, saturation, shift, and blur. For each patch, each augmentation technique had a 50 % chance of being enabled, except for blur augmentation which had a 10 % chance. The network used was an attention-gated U-Net (AGU-Net) (9, 12) with seven spatial levels and {16, 32, 32, 64, 64, 128, 128} filters, multiscale input, and deep supervision (9, 12). Adam (29) was used as optimizer, with an initial learning rate of 0.0005 that decreased with a factor of 0.5 for every tenth epoch without improvement (16, 17). The model was trained for 500 epochs using the Dice loss function, with an early stopping of patience 200. An epoch was defined as 160 and 40 weight updates for the training and validation sets, respectively. The following three models were compared: an AGU-Net trained on TMAs only without data augmentation (model I), an AGU-Net trained on TMAs only with data augmentation (model II), and an AGU-Net trained on TMAs and WSIs with data augmentation (model III). The three models were evaluated on the validation set and on both test set 1 and 2.

The quantitative segmentation performance was evaluated on TMA core level, using the Dice similarity coefficient and pixel-wise precision and recall, disregarding the background class. Each HE/CK TMA core pair was saved on disk at x20 magnification, corresponding to 0.3448 µm/pixel. For each TMA, overlapping inference of 30 % was performed using pyFAST, resulting in segmentation predictions on TMA-level used for evaluation. The metrics were computed for the three classes on all TMA cores. Metrics were also computed exclusively for cores where the respective classes were present in either the ground truth or in the prediction, and exclusively for cores where the respective classes were present in the ground truth. In cases where the denominator was zero, the metric score for that core was set to one. This will be the case for Dice scores if the prediction and the ground truth are both zero. The best performing model was then converted to the Open Neural Network Exchange (ONNX) format for deployment in FastPathology. Dice scores were also calculated separately for different histological subtypes and for each histological grade.

An additional TMA slide was segmented by a preliminary model in FastPathology at x10 magnification. Segmentations in 26 cores (9 patients) were adjusted manually by two pathologists in QuPath. These segmentations were then used as ground truth for a new quantitative evaluation of the final model. Dice scores were calculated for the three classes on TMA core level for all cores.

A qualitative evaluation of the segmentations of the final model was also performed by two pathologists through manual inspection of all TMA cores in the test set slides (test set 1 and test set 2) in QuPath. The TMA slides were segmented by the final model in FastPathology at x10 magnification and exported to QuPath. Each case was assigned a score between zero and five, similar as done by Valkonen *et al.* (18). The scoring system is described in detail in Table 1.

| Table 1: Qualitative evaluation scoring system. | | | | | |
| --- | --- | --- | --- | --- | --- |
| 0 | 1 | 2 | 3 | 4 | 5 |
| Respective class is not present in image or in segmentation, or image not suitable for analysis. | Completely wrong segmentation of respective class, only FP or FN segmentations. | Many cells are incorrectly segmented, either FP or FN. | The majority of the cells are correctly segmented, but some FP or FN. | The majority of the cells are correctly segmented, few FP and few FN. | All or almost all cells are correctly segmented, few FP and few FN. |
| Criteria for each score (0-5) for the qualitative assessment of segmentation masks. Abbreviations: FP = false positives, FN = false negatives. | | | | | |

The implementation was done in Python 3.8, with TensorFlow v2.10.0 for implementation and training of the AGU-Net model. Scikit-image v0.18.3 was used for image-to-image registration using phase cross-correlation (30). WSI processing and TMA extraction were performed with pyFAST v4.7.0 (19, 20). The experiments were performed on an Intel Xeon Gold 6230 @2.10GHz central processing unit (CPU) with 256 GB RAM, and an NVIDIA Quadro RTX 6000 dedicated graphics processing unit (GPU). The source code to reproduce the experiments is made openly available at https://github.com/AICAN-Research/breast-epithelium-segmentation

## Results

Characteristics of the five cohorts are described in Table 2. The proportion of the histological subtype invasive carcinoma NST varied between 67-89 %, and the proportion of lobular carcinoma varied between 2-24 % in the five cohorts. The proportion of histologic grade 1, 2, 3 in the five cohorts varied between 10-41 %, 43-61 %, and 16-43 %, respectively.

Table 2: Characteristics of the five cohorts. Number of patients, and proportions in parentheses, for each cohort are presented.

|  | Cohort | | | | | |
| --- | --- | --- | --- | --- | --- | --- |
|  | BCS-1 | BCS-2 | BCS-3 | HUS-BC-1 | HPA-BC | Total |
| Number of patients included | 136 | 133 | 137 | 433 | 21 |  |
| Number of TMAs | 4 | 4 | 4 | 12 | 4 | 28 |
| Number of TMA cores | 307 | 296 | 305 | 1145 | 56 | 2109 |
| Number of TMA cores/patient | 1-3 | 1-3 | 1-3 | 1-6 | 1-8 |  |
| Histologic subtype N (%) | | | | | | |
| Invasive carcinoma NST | 100 (73.5) | 104 (78.2) | 122 (89.1) | 359 (82.9) | 14 (66.7) | 699 (81.3) |
| Lobular | 14 (10.3) | 15 (11.3) | 3 (2.2) | 48 (11.1) | 5 (23.8) | 85 (9.9) |
| Other | 22 (16.2) | 14 (10.5) | 12 (8.8) | 23 (5.3) | 0 (0.0) | 71 (8.3) |
| Unknown | 0 (0.0) | 0 (0.0) | 0 (0.0) | 3 (0.7) | 2 (9.5) | 5 (0.6) |
| Total | 136 (100.0) | 133 (100.0) | 137 (100.0) | 433 (100.0) | 21 (100.0) | 860 (100.0) |
| Histologic grade N (%) | | | | | | |
| 1 | 13 (9.6) | 26 (19.5) | 17 (12.4) | 179 (41.3) | 4 (19.0) | 239 (27.8) |
| 2 | 83 (61.0) | 68 (51.1) | 61 (44.5) | 184 (42.5) | 9 (42.9) | 405 (47.1) |
| 3 | 40 (29.4) | 39 (29.3) | 59 (43.1) | 67 (15.5) | 6 (28.6) | 211 (24.5) |
| Unknown | 0 (0.0) | 0 (0.0) | 0 (0.0) | 3 (0.7) | 2 (9.5) | 5 (0.6) |
| Total | 136 (100.0) | 133 (100.0) | 137 (100.0) | 433 (100.0) | 21 (100.0) | 860 (100.0) |

*In TMA tissue blocks, 2-3 cores are often punched from the same tumor and placed together as a duplet or triplet in the recipient block. For some patients, a series of 2-3 TMA cores were placed in more than one tissue block, or twice in a tissue block. In this study, a "case" was defined as a series of 2-3 TMA cores from one patient, in a single tissue block. In addition, two WSIs from two patients from BCS-2 were included. Abbreviations: TMA = Tissue microarray, NST = No special type.

Table 3 shows three scores for each model on the evaluation sets. The first score (row I) is evaluated on all TMA cores, and score one is given if the denominator is zero. The second score (row II) includes cores with positive value in either ground truth or prediction. The third score (row III) only includes cores with positive values in the ground truth. In the validation set, a total of 105, 17, and 342 TMA cores included benign, *in situ* lesions, and invasive, respectively. The corresponding numbers in test set 1 were 77, 14, and 284 cores. In test set 2, a total of 13, 3, and 56 cores included benign, *in situ* lesions, and invasive, respectively. Examples of segmentations can be seen in Figure 4.

Table 3: Dice scores, precision, and recall on TMA cores.

| Model | | Training set | Evaluation set | Aug | Dice | | | Precision | | | Recall | | |
|---|---|---|---|---|---|---|---|---|---|---|---|---|---|
| | | | | | Benign | *In situ* | Invasive | Benign | *In situ* | Invasive | Benign | *In situ* | Invasive |
| 1 | I | TMA | Validation | No | 0.40±0.44 | 0.48±0.50 | 0.71±0.15 | 0.43 | 0.49 | 0.73 | 0.83 | 0.97 | 0.72 |
| | II | | | | 0.17±0.29 | 0.03±0.15 | 0.71±0.15 | 0.21 | 0.06 | 0.73 | 0.77 | 0.94 | 0.72 |
| | III | | | | 0.41±0.31 | 0.37±0.38 | 0.71±0.15 | 0.51 | 0.69 | 0.73 | 0.46 | 0.35 | 0.72 |
| 2 | I | TMA | Validation | Yes | 0.68±0.41 | 0.82±0.37 | 0.71±0.16 | 0.73 | 0.83 | 0.74 | 0.85 | 0.98 | 0.72 |
| | II | | | | 0.30±0.33 | 0.14±0.30 | 0.71±0.16 | 0.41 | 0.17 | 0.74 | 0.67 | 0.91 | 0.72 |
| | III | | | | 0.46±0.31 | 0.60±0.32 | 0.71±0.16 | 0.61 | 0.71 | 0.74 | 0.51 | 0.62 | 0.72 |
| 3 | I | TMA + WSI | Validation | Yes | 0.72±0.39 | 0.83±0.36 | 0.72±0.16 | 0.77 | 0.84 | 0.74 | 0.85 | 0.98 | 0.73 |
| | II | | | | 0.34±0.33 | 0.15±0.30 | 0.72±0.16 | 0.44 | 0.18 | 0.74 | 0.65 | 0.90 | 0.73 |
| | III | | | | 0.47±0.30 | 0.57±0.32 | 0.72±0.16 | 0.61 | 0.69 | 0.74 | 0.52 | 0.62 | 0.73 |
| 1 | I | TMA | Test 1 | No | 0.48±0.45 | 0.50±0.49 | 0.73±0.16 | 0.50 | 0.52 | 0.73 | 0.87 | 0.96 | 0.75 |
| | II | | | | 0.21±0.31 | 0.02±0.10 | 0.73±0.16 | 0.25 | 0.05 | 0.73 | 0.81 | 0.92 | 0.75 |
| | III | | | | 0.51±0.28 | 0.22±0.23 | 0.73±0.16 | 0.61 | 0.55 | 0.73 | 0.53 | 0.16 | 0.75 |
| 2 | I | TMA | Test 1 | Yes | 0.74±0.38 | 0.82±0.38 | 0.73±0.16 | 0.76 | 0.84 | 0.74 | 0.90 | 0.97 | 0.74 |
| | II | | | | 0.38±0.35 | 0.11±0.25 | 0.73±0.16 | 0.44 | 0.20 | 0.74 | 0.75 | 0.86 | 0.74 |
| | III | | | | 0.58±0.26 | 0.44±0.35 | 0.73±0.16 | 0.67 | 0.84 | 0.74 | 0.62 | 0.40 | 0.74 |
| 3 | I | TMA + WSI | Test 1 | Yes | 0.73±0.38 | 0.83±0.37 | 0.74±0.16 | 0.76 | 0.85 | 0.75 | 0.890 | 0.97 | 0.75 |
| | II | | | | 0.37±0.35 | 0.12±0.27 | 0.74±0.16 | 0.43 | 0.20 | 0.75 | 0.76 | 0.86 | 0.75 |
| | III | | | | 0.58±0.26 | 0.46±0.35 | 0.74±0.16 | 0.68 | 0.79 | 0.75 | 0.62 | 0.43 | 0.75 |
| 1 | I | TMA | Test 2 | No | 0.50±0.45 | 0.45±0.50 | 0.69±0.23 | 0.52 | 0.46 | 0.66 | 0.90 | 0.98 | 0.74 |
| | II | | | | 0.20±0.29 | 0.04±0.17 | 0.69±0.23 | 0.24 | 0.06 | 0.66 | 0.84 | 0.96 | 0.74 |
| | III | | | | 0.55±0.18 | 0.44±0.42 | 0.69±0.23 | 0.64 | 0.67 | 0.66 | 0.57 | 0.58 | 0.74 |
| 2 | I | TMA | Test 2 | Yes | 0.85±0.29 | 0.76±0.43 | 0.69±0.23 | 0.85 | 0.78 | 0.67 | 0.93 | 0.97 | 0.74 |
| | II | | | | 0.49±0.31 | 0.10±0.27 | 0.69±0.23 | 0.50 | 0.19 | 0.67 | 0.76 | 0.89 | 0.74 |
| | III | | | | 0.64±0.14 | 0.50±0.48 | 0.69±0.23 | 0.66 | 0.94 | 0.67 | 0.68 | 0.45 | 0.74 |
| 3 | I | TMA + WSI | Test 2 | Yes | 0.79±0.34 | 0.75±0.43 | 0.70±0.23 | 0.80 | 0.75 | 0.67 | 0.92 | 0.99 | 0.75 |
| | II | | | | 0.39±0.32 | 0.13±0.30 | 0.70±0.23 | 0.41 | 0.12 | 0.67 | 0.75 | 0.96 | 0.75 |
| | III | | | | 0.57±0.20 | 0.67±0.37 | 0.70±0.23 | 0.60 | 0.66 | 0.67 | 0.64 | 0.81 | 0.75 |

Results on TMA cores level. The first row (I) for each model on each evaluation set represents scores where all TMA cores in the respective set are included. The second row (II) represents scores where only TMA cores with the respective class included either in the ground truth or in the prediction. The third row (III) represents scores where only TMA cores where the respective class is present in the ground truth are included. Abbreviations: Aug = Augmentation, TMA = Tissue microarray, WSI = Whole slide image.

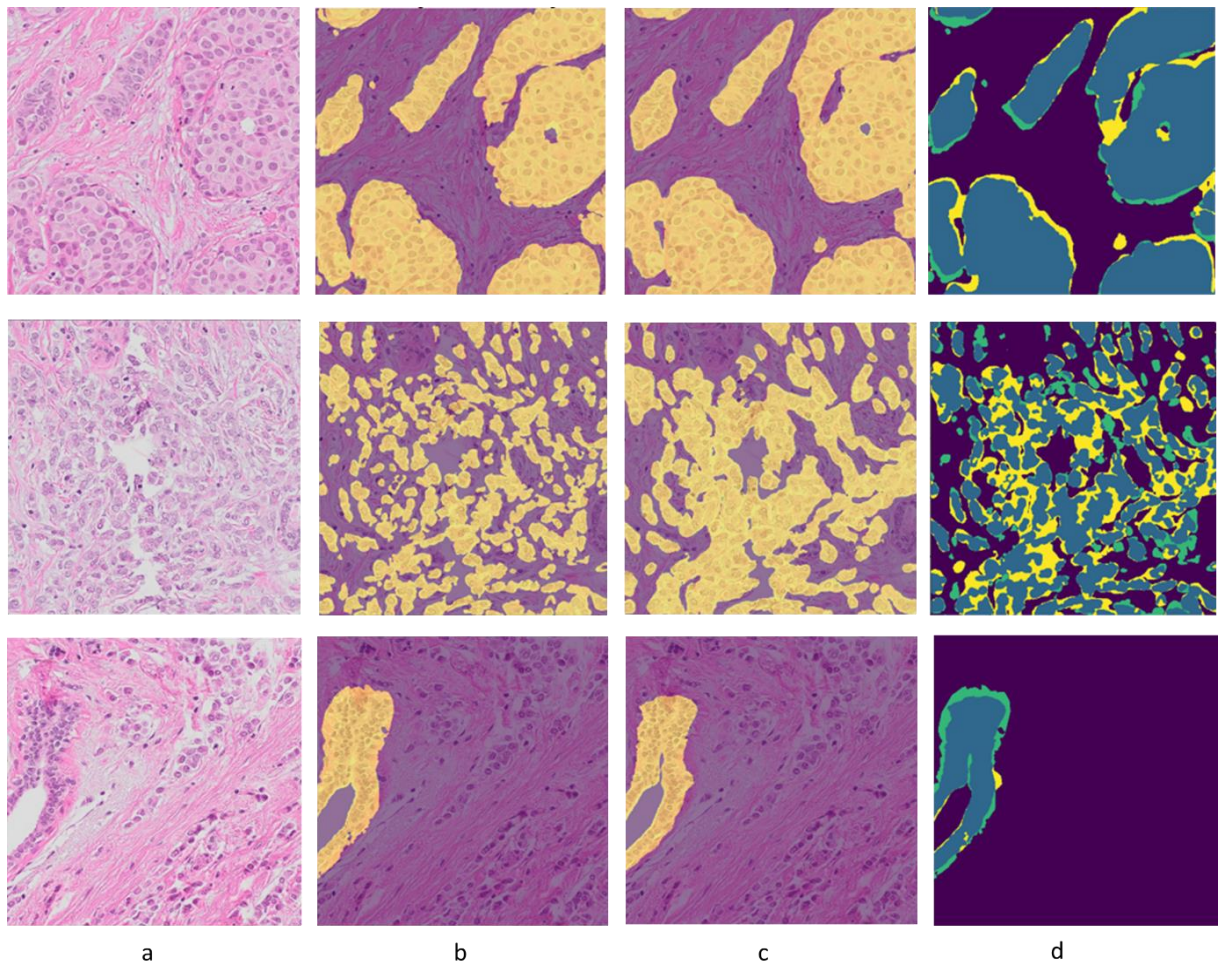

a          b          c          d

Figure 4: a) 1000 x 1000 patches of three HE images with b) corresponding ground truth, c) prediction and d) true positive (blue), false positive (yellow), and false negative (green). 1) An almost perfect segmentation of invasive cells. 2) The segmentation connects the invasive cells into larger sheets. 3) An almost perfect segmentation of parts of a benign structure. The ground truth includes the myoepithelium and fills the upper part of the lumen. Abbreviations: HE = hematoxylin and eosin.

On the TMA slide corrected by two pathologists, Dice scores of 0.56, 0.82, and 0.70 were achieved for benign, *in situ* lesions, and invasive, respectively, using model III on all cores. Precisions were 0.66, 0.82, and 0.78 and recalls were 0.82, 1.0, and 0.67 for benign, *in situ,* and invasive, respectively. When only including cores where the class was present in the ground truth, Dice scores of 0.40 and 0.70 were reached for benign and invasive, respectively. In total, six and 22 cores included benign and invasive epithelium in their ground truths, respectively. No TMA cores included *in situ* lesions.

Dice similarity coefficient was calculated separately for different histological subtypes (invasive carcinoma of no special type (NST), lobular carcinoma, and all other subtypes) and for histological grade 1-3 using model III on all TMA cores (see Table 4).

| Table 4: Dice scores on TMA core level for invasive epithelial cells on test set 1 and 2 with model III. | | |
|---|---|---|
| | Dice | |
| | Invasive | |
| Histological subtype | Test set 1 | Test set 2 |
| Invasive carcinoma NST (N=228, N = 40) | 0.75±0.14 | 0.71±0.25 |
| Lobular (N=29, N = 13) | 0.60±0.21 | 0.65±0.18 |
| Other (N=21, N = 0) | 0.73±0.23 | - |
| Histological grade | | |
| 1 (N=75, N = 11) | 0.69±0.16 | 0.68±0.29 |
| 2 (N=123, N = 15) | 0.72±0.17 | 0.74±0.13 |
| 3 (N=80, N = 27) | 0.80±0.14 | 0.67±0.26 |
| The first N (number of cores) for each subgroup corresponds to test set 1 and the second N to test set 2 for each histological subtype and grade. Abbreviations: NST = no special type. | | |

Qualitative evaluation of the TMA scores on test set 1 gave mean scores of 4.7, 3.7, 2.0, and 4.4 for all epithelium, benign epithelium, *in situ* lesions, and invasive epithelium, respectively, when excluding class zero on case level (see Table 5). On test set 2, the scores were 4.7, 3.5, 1.9, and 4.6, respectively. On test set 1, when only evaluating cores including the respective class in their ground truth, mean scores of 4.2 and 2.8 were achieved for benign and *in situ* lesions (see Table 5). An average score of 4.0 and 4.2 were reached for benign and *in situ* lesions on test set 2 when only evaluating cores including the respective class in their ground truth (see Table 5). The number of cases assigned each score can be found in Additional File 1. Model III was used to create the segmentations that were evaluated qualitatively. Examples of TMA cores with their respective score between 1-5 are shown in Figure 5-6.

| Table 5: Mean score in qualitative evaluation of segmentations. | | | | |
|---|---|---|---|---|
| | All epithelium | Benign | *In situ* | Invasive |
| Test set 1 | | | | |
| All | 4.7±0.59 (N=148) | 3.7±1.59 (N=85) | 2.0±1.30 (N=49) | 4.4±0.78 (N=148) |
| Present | | 4.2±2.0 (N=71) | 2.8±1.28 (N=27) | |
| Test set 2 | | | | |
| All | 4.7±0.61 (N=46) | 3.5±1.58 (N=19) | 1.9±1.56 (N=17) | 4.6±0.61 (N=46) |
| Present | | 4.0±1.21 (N=16) | 4.2±0.75 (N=5) | |
| Mean score (±SD) for qualitative evaluation on test set 1 and test set 2 on case level. Scores named "All" represents scores when evaluating on all cores. Scores named "Present" represents scores where only cores with the respective class in the ground truth is included. Abbreviations: SD = standard deviation. | | | | |

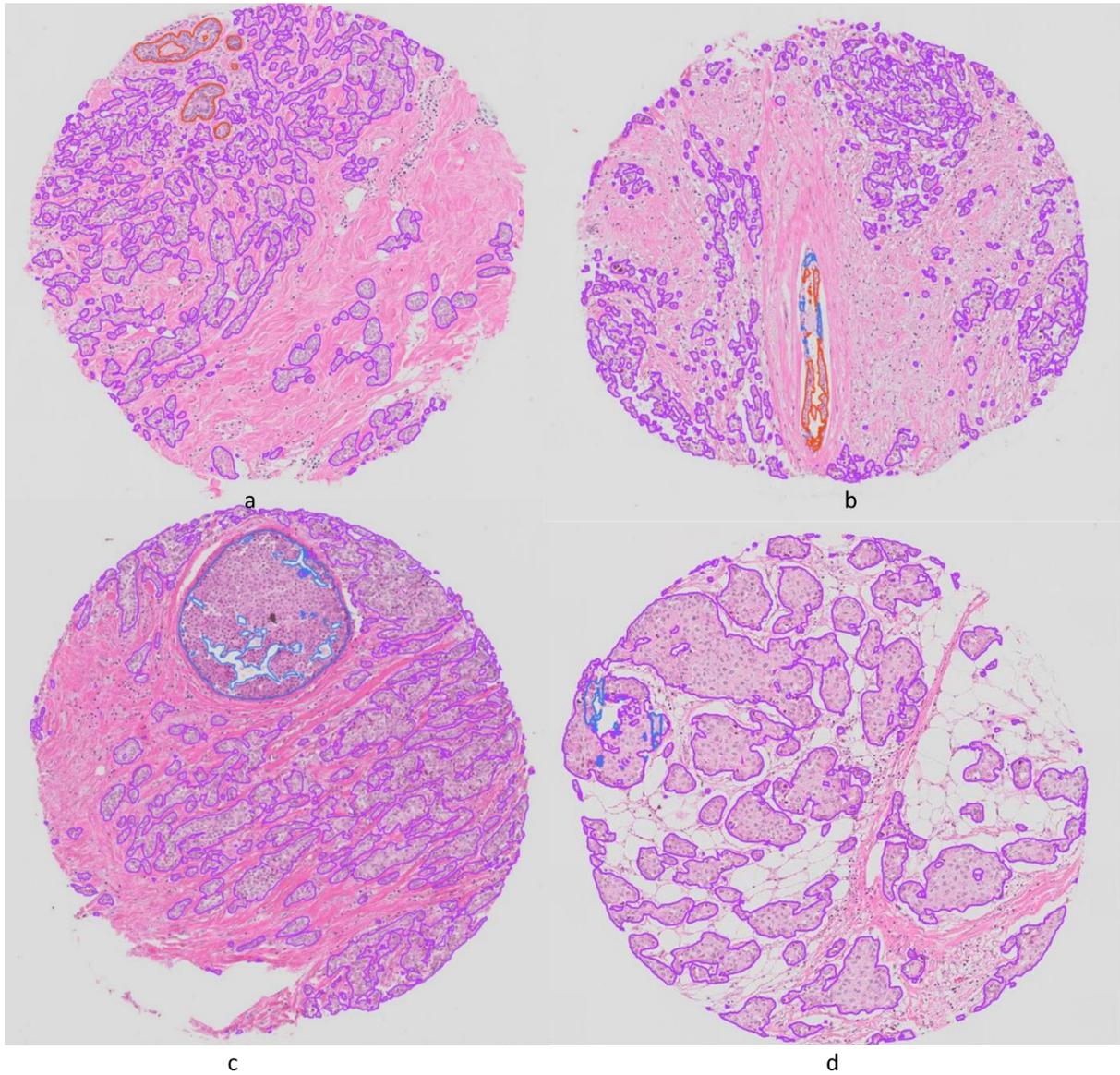

Figure 5: TMA cores and their respective qualitative scores. The scores were given on case level (two-three cores if both were present), here only one core per patient is presented. a) All epithelium score 5, benign score 5, *in situ* score 0, invasive score 5. b) All epithelium score 5, benign score 3, *in situ* score 1, invasive score 5. c) All epithelium score 5, *in situ* score 5, invasive score 4. d) All epithelium score 5, benign score 0, *in situ* score 1, invasive score 5. Red = Benign epithelium, Blue = *In situ* lesion, Purple = Invasive epithelial cells. Abbreviations: TMA = Tissue microarray.

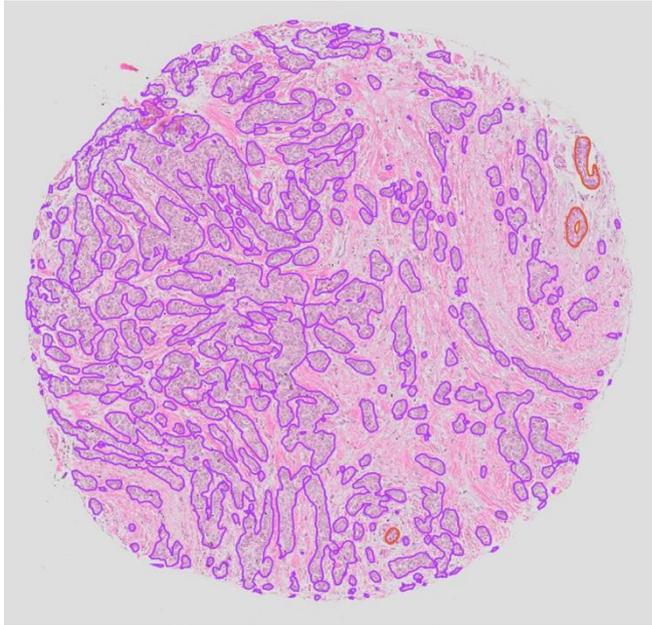

Figure 6: TMA core and its respective qualitative and quantitative scores. The qualitative scores were given on case level (two-three cores if both were present), here only one cores is presented. Qualitative scores: all epithelium score 5, benign score 5, *in situ* score 0, invasive score 5. Quantitative scores: benign and invasive Dice scores of 0.77 and 0.86, respectively. Red = benign epithelium, Purple = invasive epithelial cells. Abbreviations: TMA = Tissue microarray.

Discussion

In this study, we have developed a deep learning-based method for segmentation of benign, *in situ*, and invasive epithelial cells in HE stained breast cancer sections using IHC and pathologists' annotations to create ground truths. A new dataset comprising image pairs of HE and CK stained slides with annotations of benign and *in situ* lesions was created and used for training AI models. In qualitative evaluation, high performance was achieved for all epithelium (4.7/5) and invasive epithelial cells (4.4/5), whereas lower performances were reached for benign epithelial cells (3.7/5) and *in situ* lesions (2.0/5). In quantitative evaluation, Dice scores of 0.79, 0.75, and 0.70 were achieved for benign, *in situ*, and invasive cells, respectively, when evaluating on all TMA cores.

Other studies have used IHC for segmentation of epithelial cells in breast, colon, and prostate cancer sections (16-18). Brazdil *et al*. (17) developed an AI model for segmentation of epithelial cells using TMA slides and WSIs from breast and colon cancer. They achieved a sensitivity and specificity of 0.79 and 0.94, respectively. Their BC dataset was limited in size, including only 20 TMA cores (12 patients) and WSIs from five patients. Bulten *et al.* (16) segmented epithelial cells in sections from prostate cancer patients, separating benign and invasive epithelial cells with a myoepithelial cell marker. They achieved a Jaccard index of 0.78 and 0.83 on invasive and benign epithelium regions, respectively. In combined assessment of all epithelial cells, they achieved a Dice of 0.84 on an external test set, and a Dice of 0.88 on areas with only invasive epithelium in an internal test set. Their Dice was higher than ours for invasive epithelium. However, it is difficult to make a fair comparison to their results, as our model must correctly classify the epithelial type (benign, *in situ,* or invasive) as well as segment the cells to achieve a good score. Bulten *et al.* (16) observed a performance degradation with higher Gleason grade. We found better performance with higher histological grade on test set 1, and a poorer performance with higher histological grade on test set 2. Morphologically, high grade breast cancer cells are often pleomorphic and therefore differ more from benign epithelial cells than breast cancer cells of lower grade. Grade 1 breast cancers are often highly differentiated and may therefore be more similar to benign epithelial structures. Valkonen *et al*. (18) reached mean scores of 4.0 and 4.7 when two pathologists qualitatively scored the epithelial masks from 0 to 5. This is similar to our qualitative score of 4.7 when evaluating all epithelium. However, comparing different studies, with different methods, data material, and end points, is challenging, as the combination of many parameters and methodology choices influence the result.

Breast cancer is known for being morphologically heterogeneous (11). Grade 1 tumors generally have more ductal structures and are less pleomorphic than grade 3 tumors, and there is large variation in morphological appearance between different histological subtypes. Our model performed better on invasive carcinoma NST than on lobular carcinomas. Lobular carcinomas are often characterized by single cell growth and scant cytoplasm. These tumors are underrepresented in the dataset, and they lack certain characteristics of epithelial cells, such as the cells' tendency to cluster together. The scant cytoplasm and hence sparse CK staining of lobular invasive cells may have caused false negative CK staining. This may have led to removal of some cells during preprocessing of the CK masks, making invasive lobular cells even more underrepresented. Creating a model that performs equally well on all histological subtypes might not be possible. Artefacts like pen marks and CK marks surrounding TMA cores can also influence the ground truth, as can large shifts or broken tissue due to restaining. A non-rigid registration method could have resulted in better ground truths, and artefacts and false positive or negative staining could have been corrected either manually or by training an additional model (16). An alternative to using IHC to provide ground truth would be manual annotation of epithelial cells. This is extremely time-consuming, and not feasible if aiming to generate a large and accurate dataset. In the manual corrections that that were done on the present study for a final quantitative evaluation, the pathologists could spend up to two hours annotating a single TMA core. Two whole sections of breast cancer were added to the training data. The slides were subdivided and separated between the training and validation sets. The addition of two WSIs did not improve the model quantitatively but was chosen to increase robustness. Having patches from the same WSI in the training and validation set was not ideal. However, the two WSIs differed in the amount of benign and *in situ* lesions, and to ensure representation of both classes we allowed for using parts of the same WSI in both the training and in the validation set.

*In situ* lesions are composed of atypical epithelial cells, and thus have cellular features similar to invasive cells. They are, however, surrounded by myoepithelial cells, similar to benign epithelial structures. The poor performance on *in situ* lesions could therefore be explained by morphological similarity to both benign and invasive epithelial cells. To obtain a perfect Dice score, the model must identify and classify the epithelial structures correctly and mark the exact same cell boundary as in the ground truth. This is a challenging task. Evaluation on the slide corrected by pathologists gave lower Dice scores than evaluation of the model's segmentations against CK generated ground truths. In the manually corrected dataset, only TMA cores that were automatically excluded with pyFAST's TissueMicroArrayExtractor, or impossible to annotate manually due to very low tissue quality were removed. This could have led to the inclusion of more cores with fragile or broken tissue in this dataset than in the dataset for evaluation against CK generated ground truths, thus making the task more challenging. Furthermore, making accurate manual delineations of individual cells is a challenging task and ground truths created by CK staining could be more precise than manual annotations.

The low number of non-invasive lesions and morphological heterogeneity within these (31) could have affected the models' performance. TMAs were taken selectively from the invasive tumor region, probably affecting the amount of non-invasive epithelial tissue present. An advantage of using TMAs is, however, the inclusion of more patients. A more extensive dataset, including more patients and WSIs might still be needed to improve model performance. The low numbers of *in situ* lesions and benign epithelium give an unnaturally high Dice score when including all cores in the calculations since a score of one is given when the model correctly predicts no pixels of the given class. On the other hand, a single misclassification would more strongly influence the result of an underrepresented class.

All slides were stained and scanned at the same laboratory. The models' results were evaluated both quantitatively and qualitatively, which is important as the quantitative scores may not be representative of the segmentations due to incorrect ground truths. Metrics like Dice similarity coefficient might not be ideal to evaluate the model's performance, and a perfect Dice score may not be necessary for clinical use. A pathologist's qualitative evaluation could provide a more relevant score. The requirements of a model's performance may depend on the task for which it is used. For some clinical tasks a segmentation model needs near perfect results.

To further improve the model, one could iteratively improve the ground truth masks and consequently the model through active learning. Ground truths could be obtained by repetitive correction of annotations through multiple iterations of running the model followed by correction of the masks. Experimenting with different image levels might also improve the results. By making the final model openly available in FastPathology (https://github.com/AICAN-Research/FAST-Pathology) anyone can use it to generate segmentations on their own digitized tissue slides.

## Conclusion

The proposed method used IHC and pathologists' annotations for ground truths. The resulting segmentation model performed well in detecting epithelial cells and invasive epithelial cells in sections from breast cancer patients. However, correct classification of benign epithelial structures and *in situ* lesions was more challenging. The need for large, annotated datasets and the great morphologic heterogeneity in breast cancer represent challenging aspects of model development.

## List of abbreviations

AGU-Net = Attention-Gated U-Network

AI = Artificial Intelligence

BC = Breast Cancer

DAB = 3,3'-diaminobenzidin

CK = Cytokeratin

CPU = Central Processing Unit

EFI = Extended Focal Imaging

FN = False Negative

FP = False Positive

GPU = Graphics Processing Unit

HE = Hematoxylin and Eosin

HIER = Heat Induced Epitope Retrieval

ONNX = Open Neural Network Exchange

NST = No Special Type

TIFF = Tagged Image File Format

TMA = Tissue microarray

TP = True Positive

WSI = Whole Slide Image

## Declarations

1. Ethics approval and consent to participate: This study was approved by the Central Norway Regional Committee for Medical and Health Research Ethics (2018/2141 and 836/2009). The need for consent was waved.
2. Consent for publication: Not applicable.
3. Availability of data and materials: The dataset generated and/or analyzed during the current study are not publicly available according to the ethics approval for this study. The project code is available at https://github.com/AICAN-Research/breast-epithelium-segmentation, and an archived version is available at https://github.com/AICAN-Research/breast-epithelium-segmentation/releases/tag/v0.1.0
4. Competing interests: The authors declare that they have no competing interests.
5. Funding: The work was funded by The Liaison Committee for Education, Research, and Innovation in Central Norway [Grant Number 2018/42794], the Joint Research Committee between St. Olavs hospital and the Faculty of Medicine and Health Sciences, NTNU (FFU) [Grant Number 2019/38882 and 2021/51833], and the Clinic of Laboratory Medicine, St. Olavs hospital, Trondheim University Hospital, Trondheim, Norway. The work was also supported by grants from the Research Council of Norway through its Centres of Excellence funding scheme, project number 223250 (to LAA).
6. Authors' contributions: MH: method development, code, manuscript - writing draft, review and editing. AP: method development, code, manuscript - writing draft, review and editing. VGD: annotations, manuscript - review and editing. SMB: annotations, manuscript – review and editing. BY: staining and scanning, manuscript - review and editing. CL: Human Protein Atlas tissue processing and TMA generation, manuscript – review and editing. EW: method development, manuscript - review and editing. LAA: method development, manuscript - review and editing. IR: method development, supervision, manuscript - writing draft, review and editing. ES: method development, code, supervision, manuscript - writing draft, review and editing. MV: method development, supervision, annotations, manuscript - writing draft, review and editing.

Additional File 1:

Table 5: Qualitative results on test set 1 and 2.

| Score | All epithelium | | | | Benign | | | | *In situ* | | | | Invasive | | | |
|---|---|---|---|---|---|---|---|---|---|---|---|---|---|---|---|---|
| | Test set 1 | | Test set 2 | | Test set 1 | | Test set 2 | | Test set 1 | | Test set 2 | | Test set 1 | | Test set 2 | |
| | All | Pre | All | Pre | All | Pre | All | Pre | All | Pre | All | Pre | All | Pre | All | Pre |
| 0 | 0 | 0 | 2 | 2 | 63 | 79 | 29 | 32 | 99 | 121 | 31 | 43 | 0 | 0 | 2 | 2 |
| 1 | 0 (0.0) | 0 (0.0) | 0 (0.0) | 0 (0.0) | 18 (0.21) | 2 (0.03) | 3 (0.16) | 0 (0.0) | 26 (0.53) | 4 (0.15) | 12 (0.71) | 0 (0.0) | 0 (0.0) | 0 (0.0) | 0 (0.0) | 0 (0.0) |
| 2 | 1 (0.01) | 1 (0.01) | 1 (0.02) | 1 (0.02) | 3 (0.04) | 3 (0.04) | 3 (0.16) | 3 (0.19) | 9 (0.18) | 9 (0.33) | 0 (0.0) | 0 (0.0) | 2 (0.01) | 2 (0.01) | 0 (0.0) | 0 (0.0) |
| 3 | 8 (0.05) | 8 (0.05) | 1 (0.02) | 1 (0.02) | 8 (0.09) | 8 (0.12) | 2 (0.11) | 2 (0.11) | 7 (0.14) | 7 (0.26) | 1 (0.06) | 1 (0.2) | 21 (0.14) | 21 (0.14) | 3 (0.07) | 3 (0.07) |
| 4 | 21 (0.14) | 21 (0.14) | 7 (0.15) | 7 (0.15) | 16 (0.19) | 16 (0.23) | 3 (0.16) | 3 (0.19) | 3 (0.06) | 3 (0.11) | 2 (0.12) | 2 (0.4) | 38 (0.26) | 38 (0.26) | 13 (0.28) | 13 (0.28) |
| 5 | 118 (0.80) | 118 (0.80) | 37 (0.80) | 37 (0.80) | 40 (0.48) | 40 (0.58) | 8 (0.42) | 8 (0.5) | 4 (0.08) | 4 (0.15) | 2 (0.12) | 2 (0.4) | 87 (0.59) | 87 (0.59) | 30 (0.65) | 30 (0.65) |

Number of cases assigned to each score for each class (all epithelium, benign, *in situ*, and invasive). The scores under "Pre" represent scores where only cores with the respective class *in the ground truth* are included, otherwise the score is set to zero. The numbers in parentheses represent the percentage excluding score zero.